\UseRawInputEncoding 
\documentclass[aps,prl,reprint,superscriptaddress]{revtex4-2}
\usepackage{amssymb}
\usepackage{graphicx}
\usepackage{dcolumn}
\usepackage[version=3]{mhchem}
\usepackage{textcomp, gensymb}
\usepackage[english]{babel}
\usepackage[T1]{fontenc}
\usepackage{xcolor}

\makeatletter
\def\bbl@set@language#1{%
  \edef\languagename{%
    \ifnum\escapechar=\expandafter`\string#1\@empty
    \else\string#1\@empty\fi}%
  \@ifundefined{babel@language@alias@\languagename}{}{%
    \edef\languagename{\@nameuse{babel@language@alias@\languagename}}%
  }%
  \select@language{\languagename}%
  \expandafter\ifx\csname date\languagename\endcsname\relax\else
    \if@filesw
      \protected@write\@auxout{}{\string\select@language{\languagename}}%
      \bbl@for\bbl@tempa\BabelContentsFiles{%
        \addtocontents{\bbl@tempa}{\xstring\select@language{\languagename}}}%
      \bbl@usehooks{write}{}%
    \fi
  \fi}
\newcommand{\DeclareLanguageAlias}[2]{%
  \global\@namedef{babel@language@alias@#1}{#2}%
}
\makeatother

\DeclareLanguageAlias{en}{english}

\begin{document}
\author{Christian P. N. Tanner}
\affiliation{Department of Chemistry, University of California, Berkeley, CA 94720, USA}

\author{Vivian R. K. Wall}
\affiliation{Department of Chemistry, University of California, Berkeley, CA 94720, USA}

\author{Mumtaz Gababa}
\affiliation{Department of Chemistry, University of California, Berkeley, CA 94720, USA}

\author{Joshua Portner}
\affiliation{Department of Chemistry, James Franck Institute, and Pritzker School of Molecular Engineering, University of Chicago, Chicago, IL 60637, USA}

\author{Ahhyun Jeong}
\affiliation{Department of Chemistry, James Franck Institute, and Pritzker School of Molecular Engineering, University of Chicago, Chicago, IL 60637, USA}

\author{Matthew J. Hurley}
%\altaffiliation{Present Address: Department of Physics, Stanford University, Stanford, CA 94305, USA}
\affiliation{Department of Physics, Arizona State University, Tempe, AZ 85287, USA}

\author{Nicholas Leonard}
\affiliation{Department of Physics, Arizona State University, Tempe, AZ 85287, USA}

\author{Jonathan G. Raybin}
\affiliation{Department of Chemistry, University of California, Berkeley, CA 94720, USA}

\author{James K. Utterback}
\altaffiliation{Present Address: Sorbonne Universit\'e, CNRS, Institut des NanoSciences de Paris, INSP, 75005 Paris, France}
\affiliation{Department of Chemistry, University of California, Berkeley, CA 94720, USA}

\author{Ahyoung Kim}
\altaffiliation{Present Address: California Institute of Technology, Pasadena, CA 91125, USA}
\affiliation{Department of Chemistry, University of California, Berkeley, CA 94720, USA}

\author{Andrei Fluerasu}
\affiliation{Brookhaven National Laboratory, NSLS-II, Upton, NY 11973, USA}

\author{Yanwen Sun}
\affiliation{Linac Coherent Light Source, SLAC National Accelerator Laboratory, Menlo Park, CA 94025, USA}

\author{Johannes M\"oller}
\affiliation{
European X-ray Free-Electron Laser Facility, Holzkoppel 4, 22869 Schenefeld, Germany}

\author{Alexey Zozulya}
\affiliation{
European X-ray Free-Electron Laser Facility, Holzkoppel 4, 22869 Schenefeld, Germany}

\author{Wonhyuk Jo}
\affiliation{
European X-ray Free-Electron Laser Facility, Holzkoppel 4, 22869 Schenefeld, Germany}

\author{Felix Brausse}
\affiliation{
European X-ray Free-Electron Laser Facility, Holzkoppel 4, 22869 Schenefeld, Germany}

\author{James Wrigley}
\affiliation{
European X-ray Free-Electron Laser Facility, Holzkoppel 4, 22869 Schenefeld, Germany}

\author{Ulrike Boesenberg}
\affiliation{
European X-ray Free-Electron Laser Facility, Holzkoppel 4, 22869 Schenefeld, Germany}

\author{Jan-Etienne Pudell}
\affiliation{
European X-ray Free-Electron Laser Facility, Holzkoppel 4, 22869 Schenefeld, Germany}

\author{J\"org Hallmann}
\affiliation{
European X-ray Free-Electron Laser Facility, Holzkoppel 4, 22869 Schenefeld, Germany}

\author{Wei Lu}
\affiliation{
European X-ray Free-Electron Laser Facility, Holzkoppel 4, 22869 Schenefeld, Germany}

\author{Roman Shayduk}
\affiliation{
European X-ray Free-Electron Laser Facility, Holzkoppel 4, 22869 Schenefeld, Germany}

\author{Mohamed Youssef}
\affiliation{
European X-ray Free-Electron Laser Facility, Holzkoppel 4, 22869 Schenefeld, Germany}

\author{Anders Madsen}
\affiliation{
European X-ray Free-Electron Laser Facility, Holzkoppel 4, 22869 Schenefeld, Germany}

\author{David T. Limmer}
\affiliation{Department of Chemistry, University of California, Berkeley, CA 94720, USA}
\affiliation{Chemical Sciences and Materials Sciences Divisions, Lawrence Berkeley National Laboratory, Berkeley, CA 94720, USA}
%\affiliation{Materials Sciences Division, Lawrence Berkeley National Laboratory, Berkeley, CA 94720, USA}
\affiliation{Kavli Energy NanoSciences Institute, University of California, Berkeley, CA 94720, USA}

\author{Dmitri V. Talapin}
\affiliation{Department of Chemistry, James Franck Institute, and Pritzker School of Molecular Engineering, University of Chicago, Chicago, IL 60637, USA}
\affiliation{Center for Nanoscale Materials, Argonne National Laboratory, Argonne, IL 60517, USA}

\author{Samuel W. Teitelbaum}
\affiliation{Department of Physics, Arizona State University, Tempe, AZ 85287, USA}

\author{Naomi S. Ginsberg}
\email{nsginsberg@berkeley.edu}
\affiliation{Department of Chemistry, University of California, Berkeley, CA 94720, USA}
\affiliation{Department of Physics, University of California, Berkeley, CA 94720, USA}
\affiliation{Molecular Biophysics and Integrated Bioimaging Division, Lawrence Berkeley National Laboratory, Berkeley, CA 94720, USA}
\affiliation{Materials Sciences and Chemical Sciences Division, Lawrence Berkeley National Laboratory, Berkeley, CA 94720, USA}
\affiliation{Kavli Energy NanoSciences Institute, University of California, Berkeley, CA 94720, USA}
\affiliation{STROBE, NSF Science \& Technology Center, Berkeley, CA 94720, USA}

\title{Origins of suppressed self-diffusion of nanoscale constituents of a complex liquid}

%\begin{document}
%TC:ignore 
\begin{abstract}

Understanding and ultimately controlling the transformations and properties of  nanoscale systems, from proteins to synthetic nanomaterial assemblies, is limited by the inability to uncover their dynamics on their characteristic length and time scales. Here, we nevertheless demonstrate this ability using MHz X-ray photon correlation spectroscopy (XPCS) -- directly elucidating the characteristic microsecond-dynamics of density fluctuations of semiconductor nanocrystals (NCs), not only in a colloidal dispersion but\textcolor{black}{, more importantly,} in a liquid phase consisting of densely packed, yet mobile, NCs with no long-range order. 
We find the wavevector-dependent fluctuation rates in the liquid phase are suppressed relative to those in the colloidal phase and relative to observations of densely packed repulsive particles. We show that the suppressed rates are due to a substantial decrease in the self-diffusion of NCs, which we attribute to explicit attractive interactions. Using coarse-grained simulations, we find that the extracted shape and strength of the interparticle potential explains the stability of the liquid phase, in contrast to the gelation observed via XPCS in many other charged colloidal systems. This work opens the door to elucidating fast, condensed phase dynamics in 
complex fluids and other nanoscale soft matter, such as densely packed proteins and non-equilibrium self-assembly processes, in addition to designing microscopic strategies to avert gelation. 
\end{abstract}
%TC:endignore 

\maketitle

Equilibrium fluctuations of a system's microscopic parameters encode its nonequilibrium response to perturbations \cite{chaikin_principles_1995}. For example, a system's equilibrium density field fluctuations or individual particle velocities encode the nature of their transport. Analogous relations exist for charge, volume, and heat flow \cite{limmer2024statistical}. Knowing and manipulating how microscopic equilibrium fluctuations control these emergent nonequilibrium properties 
thus enables direct control over dynamical system properties. One important example of this connection is determining how the 
microscopic hydrodynamic, i.e., solvent-mediated, interactions of particles in a colloidal suspension, change as a function of the interactions between the particles and their concentration \cite{jones_dynamics_1991,nagele_dynamics_1996,brady_rheological_1993}. While much work has targeted understanding  hydrodynamic interactions of model systems, such as  repulsive spheres, \cite{ladd_hydrodynamic_1990,orsi_dynamics_2012,heinen_short-time_2011,westermeier_structure_2012,dallari_microsecond_2021} \textcolor{black}{and non-equilibrium systems with short-range interactions \cite{pham_glasses_2004,riest_short-time_2018,laurati_structure_2009}} \textcolor{black}{no prior study has accessed the equilibrium fluctuations of a genuine attractive colloidal liquid, nor characterized the dynamics of any complex fluid liquid on the intrinsic length- and timescales of its constituents.}

\begin{figure*}
\includegraphics[width=16cm]{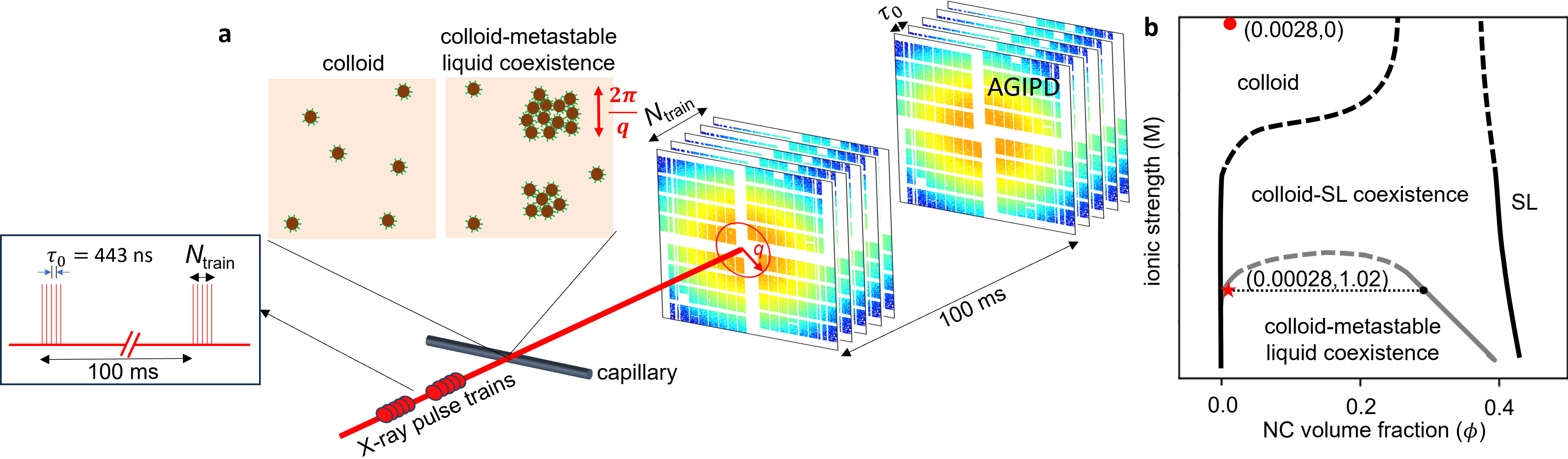}
\vspace*{-3mm}
\caption{MHz XPCS experiment overview. \textbf{a.} Ultrafast X-ray pulses are delivered to a NC solution sample  in a glass capillary.  X-rays scatter onto the AGIPD detector that is synchronized to the incident X-ray pulses. Incident X-ray pulse trains consist of $N_\text{train}$ pulses fired at 1/$\tau_0$=2.2 MHz. In this schematic, $N_\text{train}=5$, but in the experiments, $N_\text{train} = 50$. Trains are repeated at 10 Hz. Correlation analysis is performed within each pulse train and then averaged over the $\sim$100s of pulse trains within a measurement.
\textbf{b.} PbS NC volume fraction ($\phi$)--quench depth (\textcolor{black}{ionic strength}) phase diagram \textcolor{black}{adapted from Ref.  \cite{tanner_enhancing_2025}}. Solid curves indicate \textcolor{black}{previously measured} colloid-SL (black) and colloid-metastable liquid (grey) phase coexistence boundaries (SL indicates a solid phase of NCs); \textcolor{black}{dashed curves are sketched interpolation}. Red \textcolor{black}{points and associated coordinates} represent \textcolor{black}{the colloidal and liquid} state \textcolor{black}{conditions used  throughout the text, unless otherwise stated, with liquid volume fraction indicated with a black point. See SI for additional sample details.}}
\vspace*{-7mm}
\label{schematic}
\end{figure*}

Measuring the dynamics of many complex fluids is challenging due to their small lengthscales (nm) and fast timescales ($\mu$s). Typically, the translational, vibrational, or rotational dynamics of micron-scale particles are measured optically in the time domain \cite{pusey_measurement_1983,van_megen_dynamic_1991,gasser_real-space_2001,calderon_experimental_1993,savage_experimental_2009,verhaegh_fluid-fluid_1996}. Atomic systems' dynamics are measured in the frequency domain using inelastic X-ray or neutron scattering and are limited to $\lesssim$ns  \cite{sandy_hard_2018}. X-ray photon correlation spectroscopy (XPCS) is a time-domain coherent X-ray scattering technique that measures density fluctuations in reciprocal space \cite{sutton_observation_1991,fluerasu_x-ray_2005,shpyrko_x-ray_2014,sutton_review_2008,sandy_hard_2018,borsali_x-ray_2008, orsi_dynamics_2012}. MHz XPCS at the European X-ray free electron laser (XFEL) offers the ability to measure microsecond-timescale dynamics, \cite{grubel_xpcs_2007,lehmkuhler_emergence_2020} filling an important gap, provided proper accounting for X-ray-induced effects \cite{lehmkuhler_emergence_2020,reiser_resolving_2022,lehmkuhler_dynamics_2018}.

Despite the fast timescales and short lengthscales of nanoscale systems, we directly elucidate the dynamics associated with density fluctuations of few-nanometer-diameter charged semiconductor nanocrystals (NCs) as a function of their volume fraction and interaction strength using MHz XPCS. By changing the ionic strength of the dispersing solution of this complex fluid, we tune  electrostatic repulsion between the charged NCs, thereby controlling their  interaction strength. Consequently, we are able to generate both stable colloidal and liquid phases \cite{coropceanu_self-assembly_2022,tanner_enhancing_2025}. The liquid phase consists of densely packed, mobile NCs with no long-range order. 
We carefully disentangle the intrinsic NC dynamics  in the colloidal and liquid phases from  X-ray-induced effects. Together with theoretical methods, we extract  NC self-diffusion coefficients \cite{chaikin_principles_1995}  as a function of their volume fraction and interaction strength and find that the self-diffusion  of NCs in the liquid phase is suppressed relative to that of hard spheres \cite{heinen_short-time_2011} and charged spheres \cite{heinen_short-time_2011} at the same volume fraction. This suppression can be explained through a combination of hydrodynamic and explicit attractive interactions between NCs. We extract their interaction strength via direct comparison with simulations, finding a $\sim2$ $k_\textrm{B}T$ attractive interaction  explains the stability of this liquid phase. Comparing to other nanoscale and microscale complex fluids, we also infer the nanoscale interparticle interaction potential shape, which explains why these NCs avoid kinetic arrest or gelation following quenches from the colloidal phase, as compared to microscale systems. The combination of experiment, analysis, theory, and simulation enables us to determine the transport properties and \textcolor{black}{interparticle potential of this technologically relevant non-model complex fluid on its intrinsic length- and timescales, in both a dilute colloidal and a condensed liquid phase coexisting across a thermodynamic phase boundary. This work thus provides a transferable strategy for interrogating functionally relevant materials beyond model systems}, such as polymer systems \cite{nogales_x_2016,leheny_xpcs_2012} and proteins under physiological, cellular conditions \cite{moller_x-ray_2019,reiser_resolving_2022}.

To determine the interactions between NCs in complex fluid phases, we study 5.8$\pm$0.3 nm diameter PbS NCs coated with Sn$_2$S$_6^{4-}$ ligands dispersed in a mixture of N-methylformamide (NMF) and N,N-dimethylformamide (DMF) \cite{coropceanu_self-assembly_2022,jeong_colloidal_2024}. In addition to studying  colloidal-phase NCs (\textbf{Figure \ref{schematic}a} inset,  \textbf{\ref{schematic}b} upper red circles), we quench the system by adding a K$_3$AsS$_4$ salt solution that screens the electrostatic repulsion  that initially stabilizes  the highly charged NCs. Consequently, the quenched state consists of colloidal NCs in coexistence with NCs in a metastable liquid phase (\textbf{Figure \ref{schematic}a} inset,  \textbf{\ref{schematic}b} lower red circle) \cite{tanner_enhancing_2025}. To measure the structure and dynamics of NCs in the colloidal and liquid phases, we use MHz XPCS at the Materials Imaging and Dynamics (MID) instrument at the European XFEL (\textbf{Figures 1, S1}) \cite{madsen_materials_2021,tschentscher_photon_2017,decking_mhz-repetition-rate_2020}. The NC solutions scatter X-rays generated from pulse trains consisting of $N_\text{train}$ = 50 ultrafast ($\sim$50 fs-duration), spatially coherent X-ray pulses at a 2.2 MHz repetition rate onto the detector \cite{henrich_adaptive_2011} located 7 m downstream. By azimuthally integrating detector images like in \textbf{Figure 2a}, we obtain one-dimensional SAXS patterns, $I(q)$, which describe the scattered X-rays' intensity as a function of momentum transfer, $q$ (\textbf{Figure 2b}). By fitting the background-subtracted SAXS patterns to models of colloidal and liquid phases (\textbf{Figure S2}, \textbf{Methods}), we extract the NC size distribution %as well as the 
and  liquid-phase NC volume fraction --- 0.3 for the example in \textbf{Figure 2b}.

\begin{figure}[b]
\vspace*{-4mm}
\includegraphics[width=7.5cm]{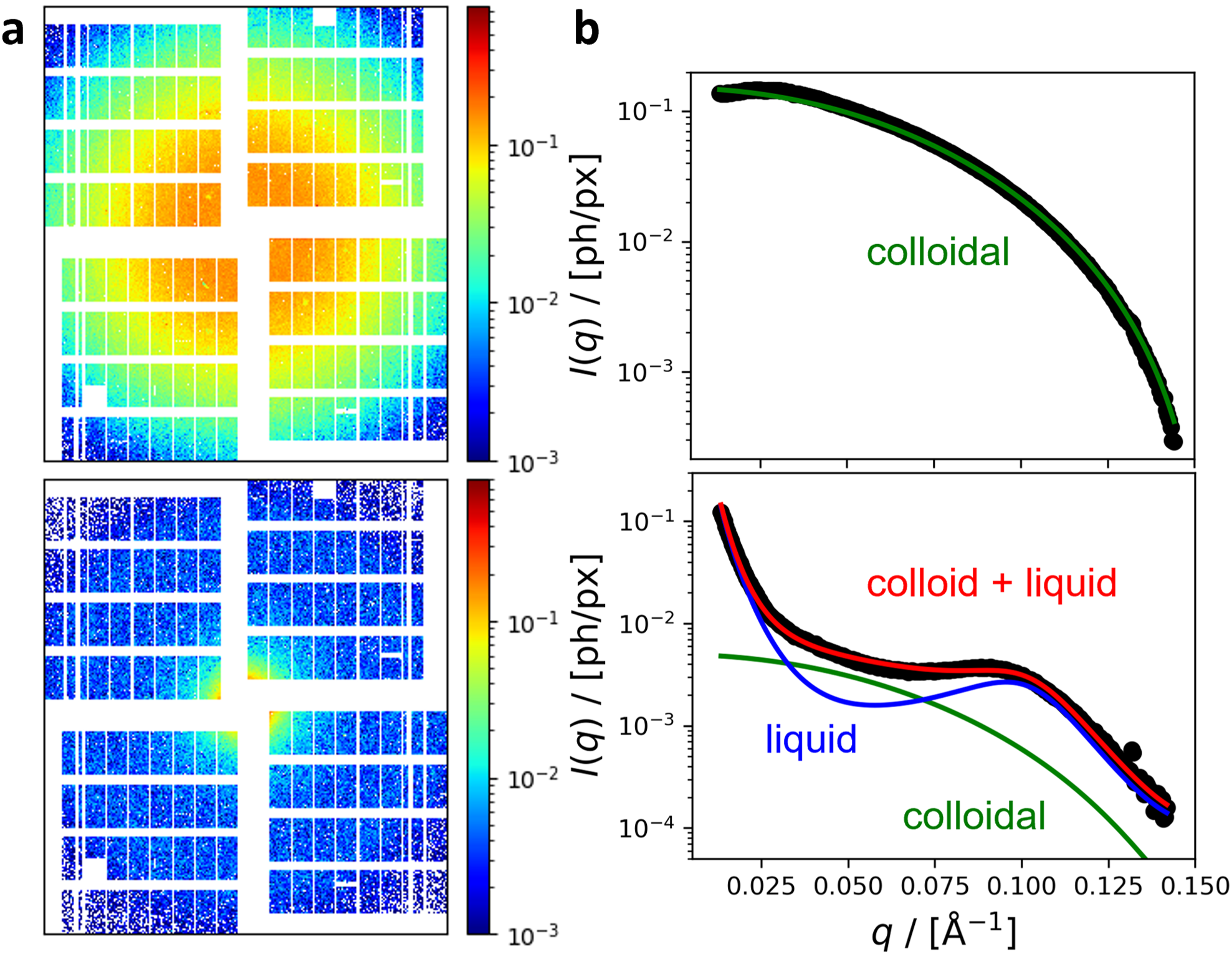}
\vspace*{-4mm}
\caption{Static scattering of NCs. \textbf{a.} Detector images for colloidal (top) and quenched (bottom) phases averaged over several pulse trains. \textbf{b.} Background-subtracted 1D SAXS patterns, $I(q)$ (units of photons per pixel), for colloidal (top) and quenched (bottom) phases (black points). Labeled solid curves indicate fits to the SAXS patterns and components.}
\vspace*{-6mm}
\label{saxs}
\end{figure}

To obtain the characteristic dynamics associated with density fluctuations of the NCs, we calculate two-time correlation functions, $C(q,t_1,t_2)$, according to  
\vspace{-2.6mm}
$$C(q,t_1,t_2)=\frac{\langle\delta I(\textbf{q},t_1)\delta I(\textbf{q},t_2)\rangle_\text{q}}{\langle I(\textbf{q},t_1) \rangle_\text{q}\langle I(\textbf{q},t_2) \rangle_\text{q}}, 
\vspace{-2.9mm}$$ 
where $\delta I(\textbf{q},t_{1})=I(\textbf{q},t_{1})-\langle I(\textbf{q},t_{1}) \rangle_\text{q}$ and $\langle ... \rangle_\text{q}$ denotes an average over $\textbf{q}$-values sharing a given magnitude of $q$. The  $C(q,t_1,t_2)$ were calculated over the 50 pulses within each pulse train and then averaged across all  trains for each experiment (\textbf{Methods}). Examples of $C(q,t_1,t_2)$ for  NCs in the colloidal and liquid phases are shown in \textbf{Figure 3a}.~\cite{Note1}. Distance along the  \textbf{Figure 3a} plots' diagonals indicates the age, $t=(t_1+t_2)/2$, of the sample within the pulse train, i.e., how many pulses the sample has sustained.  Following  $C(q,t_1,t_2)$  perpendicular to the diagonal provides the decay of the correlation function at a fixed age. The antidiagonal decay of $C(q,t_1,t_2)$ also depends on $t$, indicating that  sample dynamics change as more X-ray pulses within a  train interact with the sample. To quantify these dynamics, we calculate autocorrelation functions, $g^{(2)}\textcolor{black}{(q,\tau,\bar{t})}$, from $C(q,t_1,t_2)$\textcolor{black}{, where $\tau$ is the lag time and $\bar{t}$ is the average age of all pulse pairs used in the $g^{(2)}$ calculation} (\textbf{Methods}). 
For each phase, autocorrelation functions from $q=0.016$ (mauve) to 
0.030 
(purple) $\textrm{\AA}^{-1}$ using averages over 10  pulse \textcolor{black}{pairs per lag time} are shown as points  in \textbf{Figures 3b\textcolor{black}{, S4a}}. The autocorrelation functions of the colloidal and liquid phases decay faster at higher $q$  and decay more slowly in the liquid phase than the colloidal phase. 
%\vspace*{-5mm}
\begin{figure*}
\includegraphics[width=12cm]{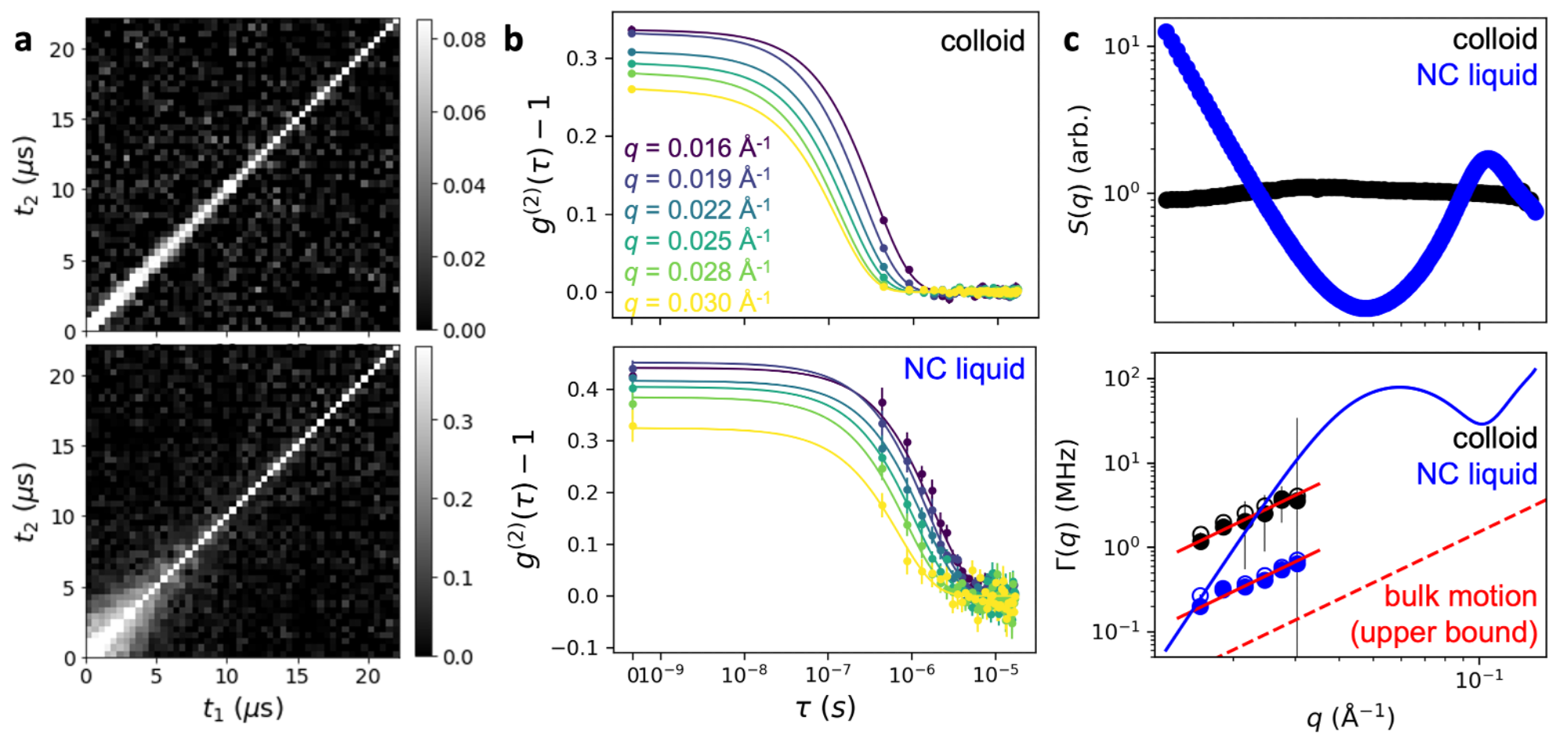}
\vspace*{-5mm}
\caption{Microsecond dynamics associated with density fluctuations of NCs in colloidal and liquid phases. \textbf{a.} Two-time correlation functions, $C(q,t_1,t_2)$, for colloidal (top) and liquid (bottom) phases at $q=0.016$ $\text{\AA}^{-1}$. \textbf{b.} \textcolor{black}{Autocorrelation functions $g^{(2)}(q,\tau,\bar{t}=\textcolor{black}{6.42}$ $\mu\text{s})$ as  increasing functions of $q$ for  colloidal (top) and liquid (bottom) phases (\textcolor{black}{colored points}), plotted on a linear scale from $\tau=0-10^{-9}$ s and log scale for $\tau>10^{-9}$ s.} \textcolor{black}{Vertical error bars represent the standard error of the mean on the 10 pulse pairs averaged for each unique lag time.} Solid curves are exponential fits. \textbf{c.} Colloidal and liquid phase structure factors, $S^c(q)$ and $S^\ell(q)$ (top), and decorrelation rates, $\Gamma(q)$ (bottom), for  colloidal (black) and liquid (blue) phases. Open circles (lower panel) indicate $\Gamma(q,\bar{t}=\textcolor{black}{6.42}$ $\mu\text{s})$;  solid circles indicate $\Gamma(q,\bar{t}\rightarrow0)$. Solid red curves are fits to $\Gamma(q)=\langle D_\text{eff}^i\rangle q^2$. Dashed red curve represents  decorrelation rate expected for bulk motion of $\sim$90-nm liquid droplets. Solid blue curve is  expected rate for  liquid phase NCs in  absence of attractive and hydrodynamic interactions ($\Gamma(q)=D_0q^2/S^\ell(q)$).}
\vspace*{-7mm}
\label{corr}
\end{figure*}

To extract the rate at which density fluctuations decorrelate, we fit an exponential model \textcolor{black}{\cite{madsen_beyond_2010,dallari_real_2024,Girelli_2025}}, $g^{(2)}(q,\tau,\bar{t})=1+\beta(q)\exp[-2\Gamma(q,\bar{t})\tau]$ (curves in \textbf{Figure 3b}), to the autocorrelation functions and obtain $\Gamma(q,\bar{t})$ (\textbf{Methods}, \textbf{Figure S3}). 
The autocorrelation functions vary as a function of $\bar{t}$ (\textbf{Figure S4\textcolor{black}{b}}). To obtain the decorrelation rate associated with the unperturbed NC dynamics, we extrapolate $\Gamma(q,\bar{t})$ to $\bar{t}\rightarrow0$ (\textbf{Figure S4\textcolor{black}{c}}). The  
rates, $\Gamma(q,\bar{t}=\textcolor{black}{6.42}$ $\mu\text{s})$ (open circles) and $\Gamma(q,\bar{t}\rightarrow0)$ (closed circles), are shown in \textbf{Figure 3c} in relation to the static structure factors, $S^c(q)$ and $S^\ell(q)$, which respectively describe the modulation in the measured scattered X-ray intensity due to the spatial arrangement of NCs in the colloidal and liquid phases (\textbf{Methods}, \textbf{Figure S5}). Here, $S^c(q)$$\sim$1 (\textbf{Figure 3c}, black points), indicating the colloidal-phase NCs are too far apart to interact 
\cite{jeong_colloidal_2024}. The liquid-phase structure factor, $S^\ell(q)$, has a distinct peak at $q$$\sim$0.1 $\text{\AA}^{-1}$, which corresponds to a $\sim$6.3 nm  \textcolor{black}{separation, presumably the} distance between neighboring NCs (\textbf{Figure 3c}, blue points\textcolor{black}{; SI}). The $\sim$1.7 peak height satisfies the Hansen-Verlet criterion for a liquid \cite{hansen_phase_1969}. Decorrelation rates $\Gamma(q,\bar{t}\rightarrow0)$ and $\Gamma(q,\bar{t}=\textcolor{black}{6.42}$ $\mu\text{s})$ are $\sim$1 order of magnitude larger for the colloidal phase than the liquid. 

To quantify these trends, we fit  $\langle D_\text{eff}^i\rangle q^2$ to $\Gamma(q,\bar{t}\rightarrow0)$ (\textbf{Figure 3c}, solid red curves) for $i\in\{c,\ell\}$, where $\langle D_\text{eff}^i\rangle$ is an effective diffusivity averaged over the fit $q$-range. We find $\langle D_\text{eff}^c \rangle$ = \textcolor{black}{45} $\mu$m$^2$/s and $\langle D_\text{eff}^\ell \rangle$ = \textcolor{black}{7} $\mu$m$^2$/s. To relate these effective diffusivities to real diffusion coefficients, we first relate $D_\text{eff}^i(q)$ to the hydrodynamic functions, $H^i(q)$, via $D_\text{eff}^i(q) = D_0H^i(q)/S^i(q)$ \cite{pusey_dynamics_1975,nagele_dynamics_1996}. Here, $D_\text{eff}^i(q)=\Gamma(q,\bar{t}\rightarrow0)/q^2$, $D_0$ is the self-diffusion coefficient of dilute NCs, and $H^i(q)$ encodes the direct, as well as the hydrodynamic, i.e., solvent-mediated, interactions between NCs. It is a many-body function that depends on all NCs in the system (\textbf{Methods}). To extract $H^i(q)$, we use the measured colloidal  and liquid  structure factors $S^i(q)$, $D_\text{eff}^i(q)$, and $D_0$ obtained from dynamic light scattering (DLS) measurements at a NC volume fraction of 0.0002 (\textbf{Figure S6}). There is minimal-to-no evidence of interactions between  NCs in the colloidal phase, as $H^c(q)\sim$ 1 (\textbf{Figure 4a}, black  points). The values obtained for $H^\ell(q)$ for liquid-phase  NCs (\textbf{Figure 4a} blue points) are, however, all below 1.

\begin{figure}
\includegraphics[width=8.4cm]{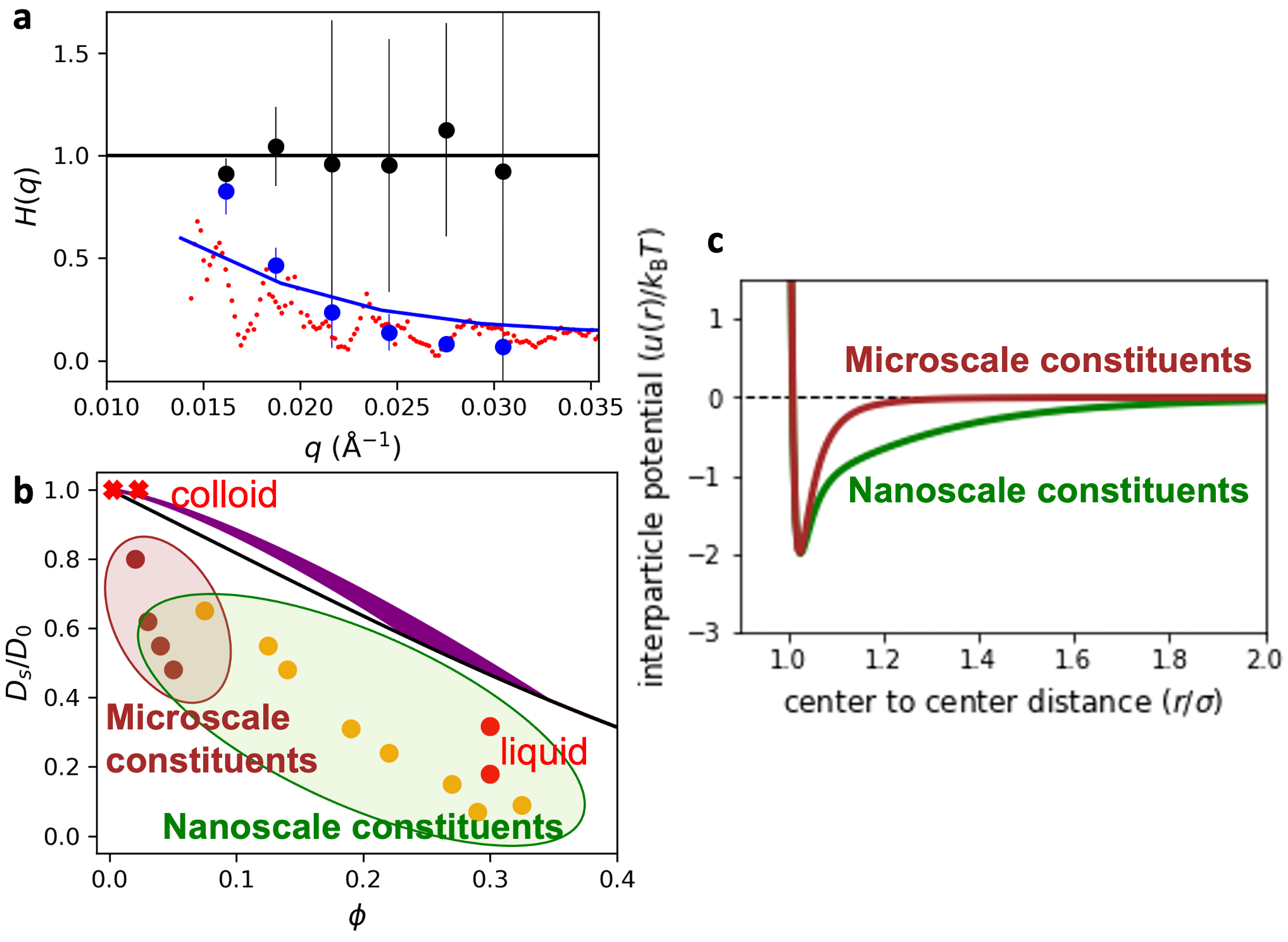}
\vspace*{-5mm}
\caption{Microscopic fluctuations of NCs as a function of  density and interaction strength. \textbf{a.} Hydrodynamic functions, $H^i(q)$, for NCs in  colloidal (black points) and liquid (blue points) phases. Solid curves are  $H^i(q)$ predictions based on  $\delta\gamma$-method  with a modified self-diffusion coefficient for  the liquid phase NCs (see \textbf{Methods}). Red points are  results of Brownian dynamics simulations of NCs interacting with a Lennard-Jones potential with  well depth $\epsilon=2$ $k_\textrm{B}T$ (oscillations arise from simulated liquid droplet's finite size). \textbf{b.} NC self-diffusion coefficient  in  colloidal phase at two  volume fractions (red x$'$s) \textcolor{black}{specified in the SI methods} and  in  liquid phase (red circles) extracted via  $\delta\gamma$-method ($D_s^\ell/D_0$=\textcolor{black}{0.31}) and via comparison to simulation ($D_s^\ell/D_0$=0.18). Solid curves are  $D_s^c/D_0$--volume fraction relationships for charged spheres \cite{heinen_short-time_2011} (purple) and hard spheres \cite{heinen_short-time_2011} (black). Gold and brown points are results from protein experiments \cite{roosen-runge_protein_2011} and micron-scale colloid-polymer mixtures \cite{seefeldt_self-diffusion_2003}, respectively \textbf{c.} Schematic nanoscale and microscale interparticle potentials  as  function of particle center-to-center distance, normalized to  particle size $\sigma$.}
\vspace*{-9mm}
\label{HI}
\end{figure}

Despite the influence of the extremely bright X-ray pulses on the sample, we  disentangle the X-ray induced effects from the intrinsic NC dynamics. \textcolor{black}{While we cannot rule out beam-induced effects on the dynamics that occur without a structural signature, the extracted decorrelation rates  (\textbf{Figure 3c}) are not due to  
NCs exiting the X-ray probe volume via convection or evaporation 
since those are $q$-independent contributions that occur on longer timescales (SI "Time-dependent heating" and \textbf{Figure S7})}. In addition, while the decorrelation rates increase with $\bar{t}$, we follow the process developed 
by Lehmk\"uhler et al. \cite{lehmkuhler_emergence_2020} 
(see \textbf{Methods}, \textbf{Figures S8, S9})  
to approximate  decorrelation rates in the absence of this temperature increase via extrapolation. Furthermore, using the $\langle D_\text{eff}^c \rangle$ = \textcolor{black}{45} $\mu\text{m}^2/$s obtained for the colloidal-phase NCs, the hydrodynamic diameter predicted through the Einstein-Smoluchowski relation is \textcolor{black}{7.7} nm. This value \textcolor{black}{is close to that measured via}
DLS (\textbf{Figure S6}). Finally, we estimate the size of individual liquid droplets to be  $\sim$30$\times$ larger than  individual NCs (\textbf{Figure S10}). Thus, the measured liquid-phase decorrelation rates  are much too fast to arise from bulk liquid droplet motion  (\textbf{Figure 3c}, dashed red curve).

Having determined the effective diffusion coefficient and hydrodynamic functions of NCs in the colloidal and liquid phases, we consider the underlying interactions consistent with the observed dynamics. The ratio of the self-diffusion coefficient, 
$D_s^i$, to $D_0$ is given by $H^i(q\rightarrow\infty)$. 
For colloidal-phase NCs, we found  $H^c(q)\sim1$ at all $q$, indicating that $D_s^c/D_0=1$ (\textbf{Figure 4b} red x's), which fall on the expected trend for charged spheres (\textbf{Figure 4b} purple curve) \cite{heinen_short-time_2011}. Since our experimental data and analysis do not sufficiently approach $q\rightarrow\infty$ to extract $D_s^\ell/D_0$, we  consider two complementary approaches. The first is  the $\delta\gamma$-method of Beenakur and Mazur \cite{beenakker_self-diffusion_1983,beenakker_diffusion_1984} (\textbf{Figure 4a} solid curves, see \textbf{Methods}). Within  our measured $q$-range  the shape of $H^\ell(q)$ is reasonably well-described by the $\delta\gamma$-method, but  a value of $D_s^\ell/D_0$=\textcolor{black}{0.31} must be included to fit the data (\textbf{Methods}). For the second approach, we performed Brownian dynamics simulations of NCs in a liquid phase interacting with various well depths $\epsilon$ and extracted $H^\ell(q)$ (\textbf{Figure 4a} red points, \textbf{Figure S11}, \textbf{Methods}). We find the extracted $H^\ell(q)$ from  simulations using $\epsilon\sim 2\textcolor{black}{\pm0.5}$ $k_\textrm{B}T$ agrees with the experimentally-determined $H^\ell(q)$ \textcolor{black}{(\textbf{Figure S11})}, and  extract $D_s^\ell/D_0$=0.18 for the simulated NCs (\textbf{Figure S12}). These two values of $D_s^\ell/D_0$ (\textbf{Figure 4b} red  points) both fall below the experimentally verified values for hard spheres (\textbf{Figure 4b}, black curve) and charged spheres (\textbf{Figure 4b}, purple curve) \cite{heinen_short-time_2011}. Given that the NCs are only an order of magnitude larger than the surrounding ion and solvent molecules, the NC solvation shell fluctuations may well play a role in suppressing the NC self-diffusion, especially as compared to larger, e.g., micron-sized, particles. 
Nevertheless, given the agreement between experiment and the Brownian dynamics simulations, we postulate that explicit attractive interactions between the NCs are predominantly responsible for the suppressed $D_s^\ell/D_0$.

Importantly, the simulated $\epsilon$=2 $k_\textrm{B}T$ NC-NC interaction strength consistent with our experimental data analysis is also consistent with values predicted from previous studies on these NCs~\cite{tanner_enhancing_2025} and from simulations \cite{haxton_crystallization_2015} of coarse-grained NCs and explains the stability of this liquid phase: in contrast to other NC systems following a quench out of the colloidal state, \cite{kpoon_gelation_1995,kovalchuk_aggregation_2012} the relatively shallow interaction strengths of the present NCs prevent their arrest in a glassy or gel-like state. This stability is crucial for the self-assembly of these NCs into ordered structures for optoelectronic applications  \cite{tanner_enhancing_2025,coropceanu_self-assembly_2022,boles_self-assembly_2016,murray_self-organization_1995,shevchenko_structural_2006,smith_self-assembled_2009,bian_shape-anisotropy_2011,santos_macroscopic_2021}. In addition, the  $D_s^i/D_0$ for the NCs in the colloidal and liquid phases could be used in rate expressions \cite{tanner_enhancing_2025,debenedetti_metastable_1996,nanev_7_2015,uwaha_8_2015} to predict and ultimately control  self-assembly kinetics, which is crucial to create defect-free structures. Further corroborating the experimental findings with hydrodynamics simulations \cite{brady_stokesian_1988,banchio_accelerated_2003,phung_stokesian_1996,heinen_short-time_2011} and extending the $q$ range over which to compare our system with models for $H^i(q)$  will further refine our intuition about the microscopic source of the suppressed diffusivities.

Returning to the  suppressed NC self-diffusion,  $D_s/D_0$ trends for colloid-polymer mixtures \cite{seefeldt_self-diffusion_2003} (\textbf{Figure 4b} brown  points) and proteins \cite{roosen-runge_protein_2011} (\textbf{Figure 4b} gold points) also fall below  expected values for hard spheres and charged spheres.  While $D_s^\ell/D_0$ for our NCs  is similar to that of proteins at the same volume fraction, colloid-polymer systems are constrained to much lower $\phi$ because gelation intervenes before metastable liquid formation. This finding suggests a difference between the interactions of measured nanoscale and microscale constituents, even though  both are typically thought to have the same interparticle interactions relative to constituent size. Simulations that phenomenologically introduced a long-range attractive tail to short-range attractive potentials previously showed that this tail effectively averted the onset of gelation at low $\phi$ \cite{noro_role_1999}. Based on these and explicit calculations of interparticle interactions using Derjaguin-Landau-Verwey-Overbeek (DLVO) theory \cite{israelachvili_intermolecular_2011} and the Asakura-Oosawa model \cite{asakura_interaction_1954,asakura_interaction_1958,vrij_polymers_1976} (\textbf{Figure S13}), we propose that nanoscale interparticle interactions, $u(r)$, likely have long-range attractive tails that are absent in the interparticle interactions of microscale particles measured to-date (\textbf{Figure 4c}). This finding provides a strategy to either avoid or enhance gelation during  self-assembly of nanoscale and microscale systems, provided the shape and depth of constituents' interparticle interactions can be sufficiently tuned. 
By extending MHz XPCS \textcolor{black}{to non-model complex fluids across} constituent characteristic length- and timescales, this work  paves the way to more generally \textcolor{black}{elucidating microscopic dynamics and} interactions in a wide range of complex fluids. These could be treated  directly or by using systems like the NCs in this work  as models for more damage-prone complex fluids, such as biomolecular condensates, whose high-resolution fluctuations are elusive to diffraction-limited optical microscopy.

\vspace*{-9mm}

\section*{Acknowledgments}
\vspace*{-5mm}
We thank A. Omar and E. Weiner for discussions. This work was supported by the Office of Basic Energy Sciences (BES), US Department of Energy (DOE) (award DE-SC0019375). We acknowledge the European XFEL in Schenefeld, Germany, for provision of XFEL 
beamtime at Scientific Instrument MID (Materials Imaging and Dynamics) under proposal 3049. C.P.N.T. and V.R.K.W. were supported by the NSF Graduate Research Fellowship. J.K.U. was supported by an Arnold O. Beckman Postdoctoral Fellowship in Chemical Sciences. D.T.L. was supported by an Alfred P. Sloan Research Fellowship. N.S.G. was supported by a David and Lucile Packard Foundation Fellowship for Science and Engineering and Camille and a Henry Dreyfus Teacher-Scholar Award. 
%TC:endignore 

\vspace*{-3mm}

\end{document}